# Spatial Scaling of Land Cover Networks

Christopher Small and Daniel Sousa

*Lamont Doherty Earth Observatory*
*Columbia University*
*Palisades, NY 10964 USA*

Spatial networks of land cover are well-described by power law rank-size distributions. Continuous field proxies for human settlements, agriculture and forest cover have similar spatial scaling properties spanning 4 to 5 orders of magnitude. Progressive segmentation of these continuous fields yields spatial networks with rank-size distributions having slopes near -1 for a wide range of thresholds. We consider a general explanation for this scaling that does not require different processes for each type of land cover. The same conditions that give rise to scale-free networks in general can produce power law distributions of component sizes for bounded spatial networks confined to a plane or surface. Progressive segmentation of a continuous field naturally results in growth of the network while the increasing perimeters of the growing components result in preferential attachment to the larger components with the longer perimeters. Progressive segmentation of two types of random continuous field results in progressions of growing spatial networks with components that achieve unity power law rank-size distributions immediately before explosive percolation occurs. This suggests that the scaling properties and uniform area distribution implied by a unity power law rank-size distribution may be a general characteristic of bounded spatial networks produced by segmenting some continuous fields. The implication is that different process-specific mechanisms may not be required to explain the rank-size distributions observed for the diversity of land cover products described here.

## Introduction

In a recent review of global land cover products it was observed that human settlements, agriculture and forest cover have similar scaling properties over a wide range of spatial scales (Small and Sousa 2015). Specifically, a variety of global land cover products, derived from different types of remotely sensed imagery, using different algorithms and assumptions, were observed to have similar scaling relationships between size and number of spatially contiguous patches of land cover. Despite the differences in land cover type and provenance, all of these products have some characteristics in common. They are all continuous fields of physical characteristics of the land surface, derived entirely or largely from remotely sensed observations. These land cover products are based on dense synoptic measurements of the physical properties of land cover. This stands in contrast to traditional maps, which are typically based on relatively sparse, qualitative observations in which the land surface characteristics are categorized into discrete classes at the time of observation. The land cover products are fundamentally different from traditional maps in the sense that they originate as synoptic continuous field measurements of physical properties that are used to produce synoptic continuous field classifications of land cover form or function. The fact that multiple independently derived depictions of three functionally disparate types of land cover exhibit very similar scaling properties at global scales suggests that these scaling properties may arise from something more fundamental than the class-specific processes normally considered responsible for Earth's land cover mosaic (e.g. urban development, cultivation, forest growth).

In this analysis, we seek a general explanation for the observed similarity of scaling that does not rely on processes specific to settlements, agriculture or forests. Because of the similarity of land cover mosaics to spatial networks, and the fact that networks are often characterized by power law scaling, we adopt the terminology and analytical approaches used in the study of networks. We begin with the observation that all three types of land cover can be represented as continuous fields that accommodate spatial heterogeneity at scales finer than the pixel resolution of the observations from which the fields are derived. This characteristic differs fundamentally from the cardinal assumption of traditional discrete thematic maps in which each geographic location is classified as one, and only one, of a set of mutually exclusive classes. We consider a more general representation of land cover as bounded spatial networks occupying a plane, or more generally a surface. In contrast to the convention of representing networks as discrete sets of nodes and links with few, if any, spatial constraints, we consider binary spatial networks that arise from discretization of continuous fields – subject to firm spatial constraints. We begin by briefly describing the different land cover products and the measurements from which they are derived, then describing the analysis by which the scaling properties are derived. The common characteristics and scaling properties of all the products are then summarized. A pair of very simple random spatial models are constructed to generate continuous fields which are used to illustrate some general properties of the spatial networks that emerge from continuous fields. From a comparison of the scaling properties of the spatial networks that emerge from these models and the global land cover products, we draw a set of general observations that may be used to understand the scaling properties of bounded spatial networks. We conclude with a discussion of the implications of common scaling properties for each class of land cover.

## Conceptual Background

In this study, we consider a specific type of spatial network obtained by discretization of a continuous spatial field. We refer to it as a bounded spatial network because it is confined to a uniform 2D lattice. The lattice is binary in the sense that each pixel (e.g. grid cell or lattice node) is an element of either the foreground network of occupied sites or the complementary background network of unoccupied sites. Using the standard network representation and terminology (see (Newman 2010)) in which nodes are connected by links and interconnected subsets of nodes comprise separate components of the network, we define the bounded spatial network so that occupied pixels are equivalent to nodes and spatially contiguous patches of pixels are equivalent to components. Shared edges of adjacent occupied pixel nodes are considered direct links. In comparison to a standard network in which the degree (number of links) each node can have is typically unbounded, each pixel node in the bounded spatial network can have at most 8 neighbors (Queen's case) by direct links so the degree of a pixel node never exceeds 8. The size distribution of components (sets of spatially connected pixel nodes) in the network replaces the degree distribution as the parameter of interest. All the pixel nodes in a component are clustered so that each pixel in the component is indirectly connected to every other pixel in the component and to no pixels outside the component. All pixels are maximally assortative in that they are only connected to the other pixels in the same component and therefore all have the same degree of connection to each other.

In comparison to scale-free networks that are characterized by power law degree distributions (Barabási and Albert 1999), the bounded spatial networks we observe have power law component size distributions. In all of the global land cover products we have analyzed, the rank-size distributions of settlements, agriculture and forest cover are all well represented by power laws in which the number of sorted components increases at the same rate at which their size decreases. Whereas scale-free networks can emerge from random networks through the combined processes of growth and preferential attachment (Barabási and Albert 1999), we show that scale-free spatial networks can also emerge from continuous fields through the same combined process of growth and preferential attachment. The process of progressive segmentation of the continuous field (explained below) naturally results in growth of the foreground network while the increase in perimeter of the growing components naturally results in preferential attachment to the largest components with the longest perimeters. In other words, the same conditions for emergence of scaling in random networks (growth & preferential attachment) can also explain emergence of scaling on bounded spatial networks. Growth is a consequence of progressive segmentation while preferential attachment is a consequence of growth within confined domain of the lattice.

The formation and evolution of a bounded spatial network involves three fundamental processes; nucleation, growth and linking. Each time a new pixel node is added to the domain, one of these three processes occurs. If the new pixel node is isolated and not adjacent to any other occupied pixel nodes we refer to the process as *nucleation*. If the new pixel node appears adjacent to a pixel node that is already part of an existing component we refer to the process as *growth*. If the new pixel node appears adjacent to two or more pixel nodes that are parts of different components we refer to the process as *linking*. If the area of the spatial domain and the component size distribution are known then the probability that addition of a new pixel node will result in either growth or nucleation can be calculated on the basis of the component perimeter distribution implied by its size distribution. Knowledge of the distribution of perimeter/area ratios of components implies knowledge of the distribution of fractal dimensions of the component distribution. As the network evolves, the distribution of component fractal dimensions evolves as does the probability of linking of components. We hypothesize that the parallel evolution of the probabilities of nucleation, growth and linking within the domain controls the emergence of scaling in bounded spatial networks. As we show with the model simulations, this combination of nucleation, growth and linking can even explain the emergence of scaling on random spatial networks – without the need to introduce land cover-specific processes.

The novelty and utility of the bounded spatial network is related to its emergence from the process of progressive segmentation of a continuous field. Segmentation refers to the process of creating two complementary sets of binary segments by imposing a threshold on a continuous field (https://en.wikipedia.org/wiki/Image_segmentation). We define the foreground network as the set of spatially contiguous segments composed of pixel nodes that exceed the threshold value and the background network as the segments composed of pixels that do not. The process of progressive segmentation over multiple thresholds can result in the processes of nucleation, growth and linking of the foreground network as new pixel node values exceed the threshold and they are added to the network. By symmetry, this results in the complementary processes of *attenuation, shrinkage* and *fragmentation* of the background network as its pixels are removed when their value exceeds the threshold. Equivalently, progressive segmentation of the field can result in attenuation, shrinkage and fragmentation of the foreground network if the order of progression of the thresholds is reversed relative to the median of the distribution of the pixel values in the field. By convention, we will refer to the foreground network as the subset of pixels that are added when the threshold approaches the median of the field distribution from above and the background network as the subset of pixels that remain below the threshold.

The utility of the continuous field, and the family of spatial networks that emerge when it is progressively segmented, is related to the generality of continuous fields for representing spatially varying characteristics of landscapes and other quantities defined on a plane or surface. In the context of land cover, the continuous field represents the general case of a distribution of fractional abundances defined over a spatial domain (e.g. (DeFries et al. 1995)). It can also represent other spatially varying scalar fields such as elevation, population density or radiance (reflected or emitted). In this study, the continuous fields represent proxies for human settlements (population density and night light brightness) and subpixel fractions of land cover (agriculture and forest) on a heterogeneous landscape. Subpixel fractions are especially useful for the ability to accommodate spatial heterogeneity across multiple scales. Progressive segmentation of a continuous field can also be seen as an analog for the temporal evolution and growth (or shrinkage) of a spatial network when the quantity represented by the field grows (or shrinks) past a fixed threshold.

**Data**

Settlements  We use two very different products to represent spatial extent of human settlements. Both products are shown superimposed at both global and regional scales in Figure 1.

*Night Light*  Stable night light composites have been derived from the Defense Meteorological Satellite Program – Operational Linescan System (DMSP-OLS) by the Earth Observation Group at the NOAA National Center for Environmental Information for every year from 1992 to 2013.  Temporal overlap of the six OLS missions allows for intercalibration of the annual mean brightness composites so that changes in the spatial extent and luminance of anthropogenic night light can be quantified globally(Elvidge et al. 2009).  The compositing process eliminates intermittent light sources like wildfires. The resulting composites give the mean brightness of each 1 km pixel as DN (digital number) values ranging from 0 (completely dark) to 65 (maximum brightness). The resulting stable lights arise primarily from human settlements (Elvidge et al. 1997a), although gas flaring and other forms of resource extraction also produce stable night light signatures in some areas (Elvidge et al. 2009).

The spatial extents of settlements and other lighted development are known to be somewhat smaller than the luminance signal detected by the OLS sensors, giving rise to a phenomenon often referred to as "overglow" (Elvidge et al. 1997b).  Because the electro-optical and atmospheric scattering processes that give rise to overglow do not change appreciably from one year to the next, and because the extent of overglow has been shown to scale with the maximum brightness of the source (Small et al. 2005), the effects of overglow are generally assumed to be a temporally stable source of uncertainty in the true spatial extent of the light sources.  Vicarious validation of night lights with higher resolution optical imagery has shown strong correspondence between growth of night lights and expansion of urban land cover (Small and Elvidge 2013).   Stable night lights are increasingly used as proxies for lighted human settlements e.g.(Sutton 2003), population density e.g. (Sutton et al. 2001), economic activity e.g. (Chen and Nordhaus 2011; Doll et al. 2006) and electrification e.g. (Doll and Pachauri 2010; Min et al. 2013). OLS night light composites are publicly available for download from http://ngdc.noaa.gov/eog/dmsp.html (access 28-11-2015).

*Population Density*  As an independent complement to night lights, we use ambient population density as a proxy for human settlements.  The LandScan ambient population density product is produced at the Oak Ridge National Laboratory of the US Department of Energy.  It is distinct from census-derived population density products in that it attempts to map ambient population distribution during the day rather than residential population density. LandScan uses a "suite of novel and dynamically adaptable" algorithms (Rose and Bright 2014) with a variety of geospatial data and remotely sensed imagery to spatially disaggregate subnational census data on global scales at a spatial resolution of 30 arc seconds (~1 km at Equator). LandScan is derived from an unsupervised classification algorithm that extracts texture based metrics from remotely sensed imagery (Vijayaraj et al. 2007). Additional information about LandScan is available at: http://web.ornl.gov/sci/landscan/index.shtml

Agriculture  We use four different products to represent spatial extent of agricultural land cover.  Each product uses different inputs but some use other global agriculture products so they are not all independent. Three of these products are shown superimposed at both global and regional scales in Figure 1.  Each product is displayed in one RGB color channel with comparable scaling so agreement among all three products is indicated by shades of gray while colors indicate disparity in the estimates of area under cultivation. The pairwise correlations among the three products shown in Figure 1 range from 0.44 to 0.57. A more detailed analysis of these products is given by Sousa and Small (2015).

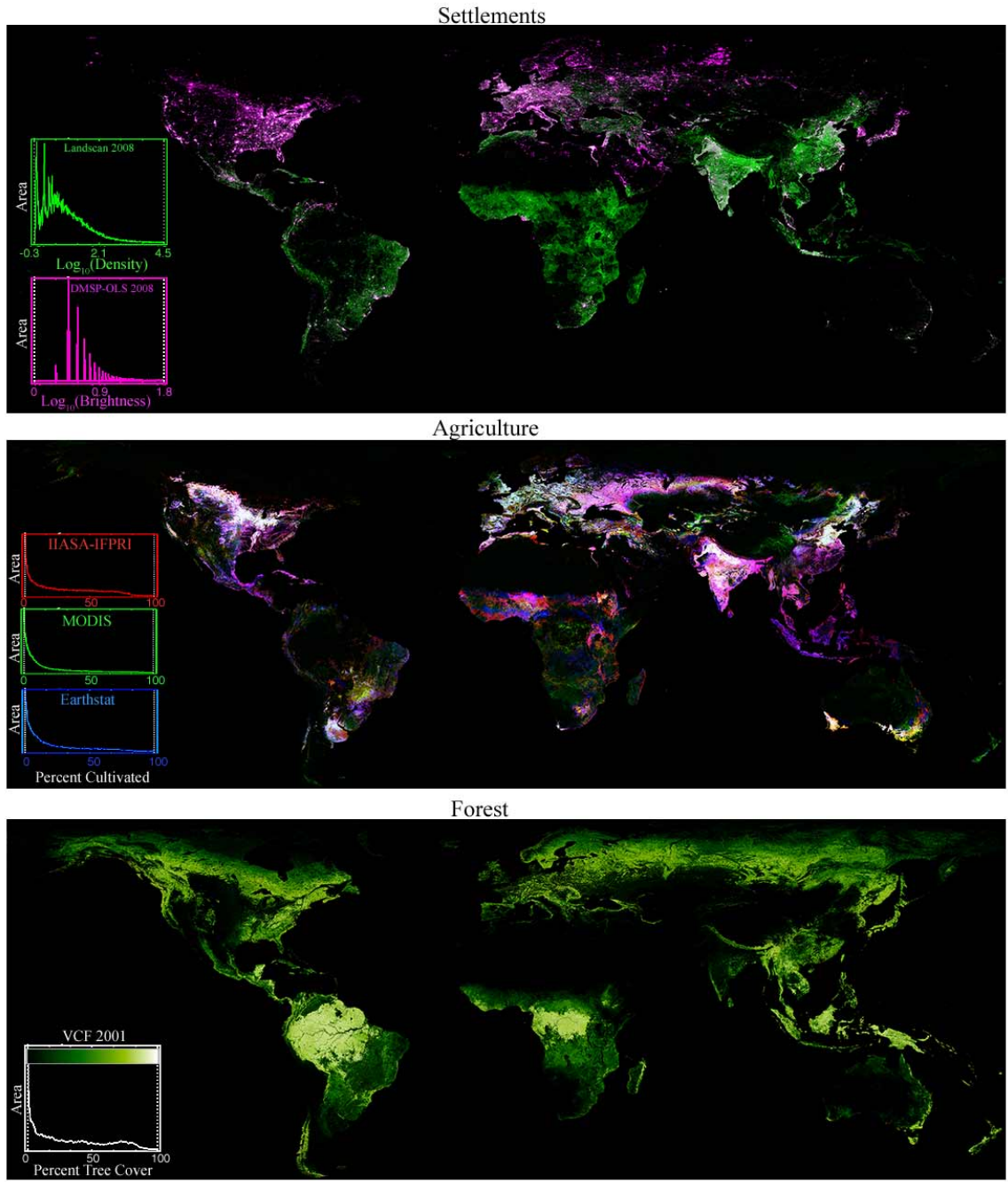

Figure 1a   Global land cover products map settlements, agriculture and forest as continuous fields at ~1 km resolution. DMSP-OLS night lights and Landscan population density products highlight the contrast between lighted urban development and high density rural populations.  Agricultural land use products show general geographic agreement but disparities in density of cultivation.  Agriculture and settlements form interspersed networks with similar geographic extent but the distribution of forest cover forms a complementary network with highest density in areas without settlements or agriculture.  Together these three spatial networks occupy areas conducive to their driving processes while oceans and barren land represent an exclusive complement.

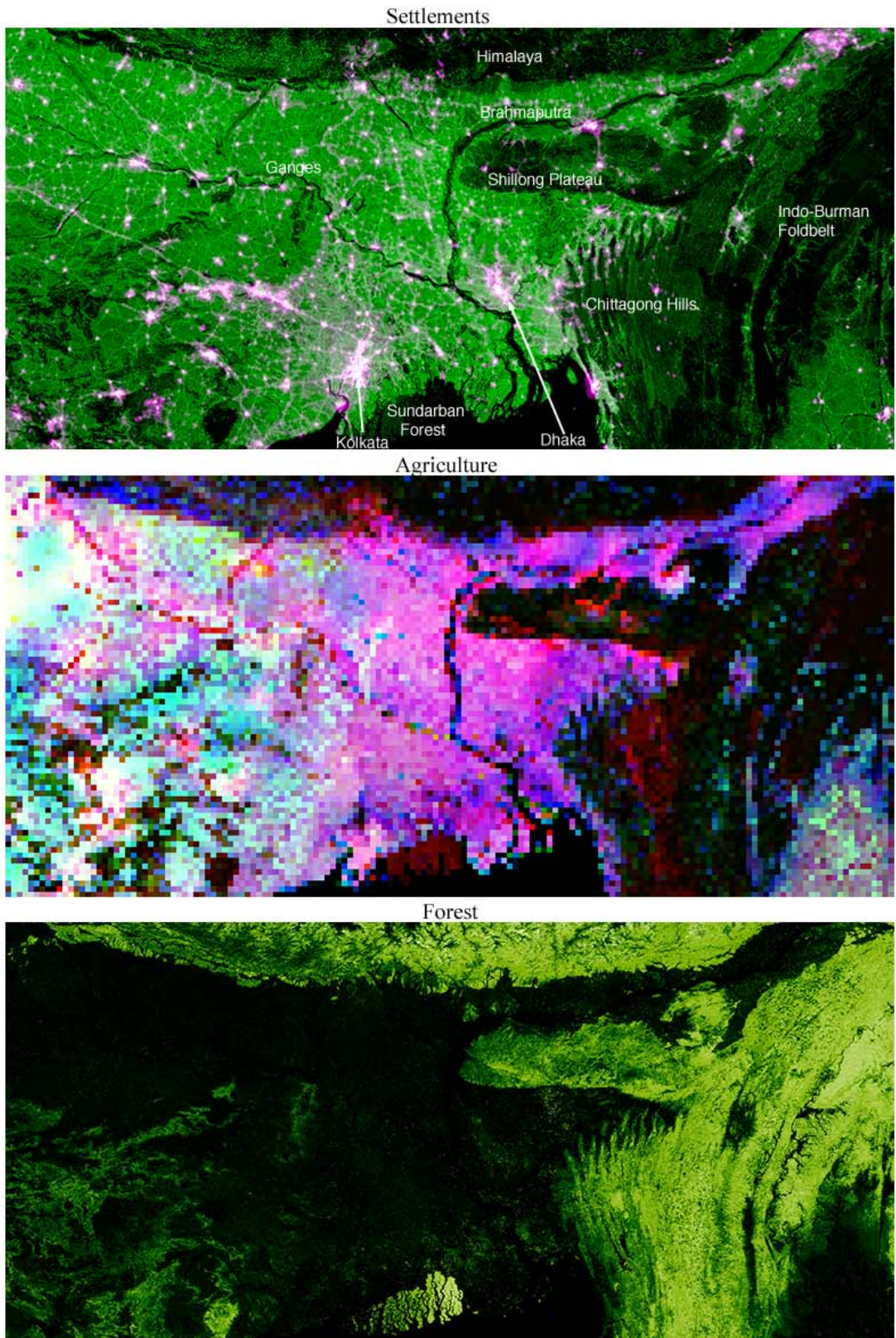

Figure 1b   Full resolution enlargements of global land cover products.  Sources and colors as in Fig. 1a.  The Ganges-Brahmaputra delta and surrounding highlands illustrate strong geographic gradients in land cover and land use with both high density agrarian populations and dense forests.

The four products used in this study are: EarthStat Global Cropland 2000 (Ramankutty et al. 2008), History Database of the Global Environment (HYDE) version 3.1 (Klein Goldewijk et al., 2010), the IIASA-IFPRI Cropland Map (Fritz et al. 2015), and the MODIS Global Cropland products (Pittman et al. 2010). All products are produced on global scales. Resolutions for the four products vary from 10 km to 500 m. All products are continuous fields represented either subpixel proportions of cropland or probability of cropland for each pixel. All of the agriculture products shown here rely on some mixture of observations based on satellite imagery and agricultural census data. The observations from the satellite imagery are generally input in the form of discrete land cover classifications. Accuracy assessments were performed in various ways for the different models. The agriculture datasets are available for download at the following URLs (access 28-08-2015):

Earthstat: http://www.earthstat.org/
HYDE 3.1: http://www.themasites.pbl.nl/tridion/en/themasites/hyde/
IIASA-IFPRI: http://www.geo-wiki.org/
MODIS Global Cropland: http://www.glad.geog.umd.edu/projects/croplands/

<u>Forests</u>
Forest cover is derived from the continuous fields Moderate Resolution Imaging Spectroradiometer (MODIS) tree cover proportion product described by (Hansen et al. 2002). Subpixel tree cover is estimated on the basis annual cycles of greening and senescence derived from time series of vegetation abundance inferred from the Normalized Difference Vegetation Index (NDVI). A regression tree algorithm is used to predict tree cover fraction at the resolution of 500 m using multitemporal MODIS imagery on the basis of the phenological signatures. The product used in this analysis gives global estimates of tree cover circa 2001, aggregated to a spatial resolution of 1 km.

## Methods

*Progressive Segmentation of Continuous Fields*
Because the land cover products we use are continuous fields, we impose fraction thresholds to create binary maps of the spatial extent and size distribution of each land cover type. To our knowledge, there are no standard definitions of settlements, agriculture or forests given as fractional land cover so we use multiple thresholds for each. By progressively segmenting the continuous field into spatially contiguous areas above and below each threshold we create a series of land cover maps that can be used to quantify threshold sensitivity. Each continuous field is projected into an equal area projection and segmented using the Queen's case in which spatial contiguity is defined on the basis of all 8 adjacent pixels. To account for spatial autocorrelation of the input imagery and avoid large numbers of spurious detections, we impose a minimum segment size of 9 pixels. A detailed description of the progressive segmentation methodology is given by (Small et al. 2011).

*Rank-Size Slope Estimation*

We quantify spatial scaling using the rank-size distribution in which the area of each spatially contiguous segment (i.e. patch) is sorted by size to yield its ordinal rank, from largest to smallest. In the context of a spatial network, the size of each contiguous segment is analogous to the size of a single network component and the rank-size plot is analogous to the component size distribution. Because the rank-size plots span several (4-6) orders of magnitude in both size and rank, we display them on log-log plots. Because the distributions generally form straight lines, we use power law exponents to quantify their slopes. While linear rank-size distributions are often assumed to be generated by power law processes, we acknowledge that it may be difficult or even impossible to conclusively eliminate other long-tailed distributions (e.g. log-normal) from consideration. Hence we use the power law merely as a tool to quantify the linearity, slope and domain of the rank-size distributions that we use to infer the spatial scaling properties of the different land cover types. To estimate the power law exponents, domain, uncertainty and validity we follow the approach given by (Clauset et al. 2009). This procedure repeatedly computes maximum likelihood estimates of the power law exponent for increasing subsets of the upper tail of the distribution and identifies a lower tail cutoff coinciding with the minimum of the Kolmogorov-Smirnov goodness-of-fit statistic. This cutoff gives an indication of the domain (range of spatial scales) over which the rank size distribution can be plausibly described with a power law.

The progressive segmentation methodology and resulting spatial networks and rank-size distributions are illustrated using night lights in Figure 2. The upper two rows show a spatial subset of the 2008 night light composite for the Midwestern United States (top left) and five successive segmentations with decreasing DN thresholds. While the highest threshold (DN >30) captures only the cores of the largest cities (e.g. Dallas) and the brightest parts of the smaller cities, lower thresholds resolve the dimmer peripheries of the larger cities and increasingly smaller towns. Progressively lowering the brightness threshold increases both the size and number of segments resolved. As adjacent segments grow and extend through their less brightly lighted peripheries, segments begin to connect to form larger spatial networks of settlements. As the threshold approaches the median of the brightness distribution, increasing numbers of segments are interconnected through the dimmer interstitial lights that exceed the falling threshold. Eventually the threshold falls below the median level of background luminance and many network components connect into much larger regional networks in a process analogous to explosive percolation. This is illustrated in the lower part of Figure 2 as a giant component forms in the eastern United States and Canada while other large components form in California and central Mexico. For the three lowest thresholds shown, each new threshold approximately doubles the total area of the network – but the slope of the rank-size distribution remains close to -1. The lower tail cutoff giving the best fit to the power law spans three to four orders of magnitude in segment size.

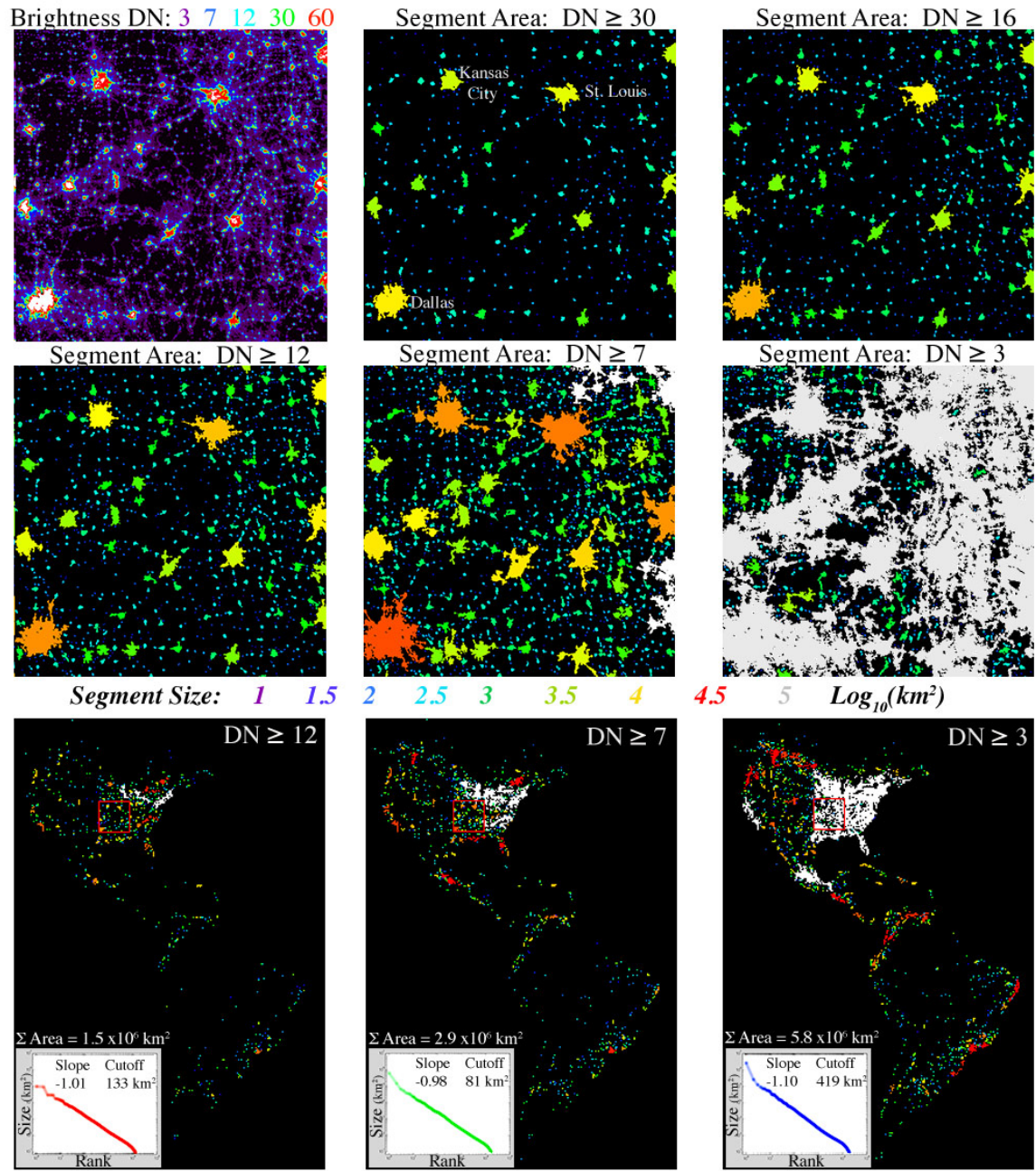

Figure 2  Progressive segmentation of night light brightness to obtain settlement size distributions. Top two rows show enlargement of midwestern USA.  Lowering DN threshold increases segment number and connectivity as smaller and dimmer lights are resolved.  The lowest threshold (DN > 3) triggers explosive percolation as several large network components connect simultaneously to form a giant component in the eastern US and Canada. Each threshold yields a scale-free distribution of spatially contiguous segments spanning 4 to 5 orders of magnitude in size. Rank-size plots (bottom insets) have consistent slopes near -1 even as total segment area doubles for each lower threshold. Progressive segmentation is analogous to a growth process that increases area and connectivity.

The progressive segmentation and rank-size power law fitting was applied to each land cover product using multiple thresholds.  Comparison of network structure and scaling using multiple thresholds gives insight into the threshold-scaling dependence of each product, while avoiding the need to rely on ad hoc definitions of what constitutes a settlement, farm or forest.  Global distributions are produced by first applying the progressive segmentation to geographic subsets and then combining the regional rank-size distributions.  The geographic subsets were chosen to avoid breaking any large spatially contiguous segments.  Using geographic subsets also allows us to compare continent-scale distributions to global distributions to evaluate their consistency.

The 21 year archive of DMSP-OLS night lights provides the opportunity to quantify the evolution of the network in places that have experienced high rates of urban development.  We apply the progressive segmentation procedure to intercalibrated night light composites for China and India in 1994, 2004 and 2013 and compare the changing spatial structure of the networks and their rank-size distributions.  An intermediate threshold of $DN \geq 8$ is used in each case, but comparable results are obtained for other thresholds below the distribution median where percolation begins.  We use 1994 as the initial year because its global brightness distribution is more consistent with subsequent years than the 1992 or 1993 composites.  As the source of this difference might be related to imaging conditions (e.g. number of cloud-free acquisitions) in these years, rather than actual differences in the light sources, we prefer the more conservative change estimates that should result from the more consistent distributions.

*Random Spatial Network Simulation*
To provide a context for the spatial networks obtained from progressive segmentation of the land cover products, we simulate the formation and evolution of random spatial networks generated by ballistic deposition processes (Barabási and Stanley 1995).  The simplest model is based on random additive deposition in which each deposition event adds a value of 1 to the randomly selected pixel.  A random multiplicative model is similarly generated but with each deposition event multiplying the pixel total by a constant (1.05).   A third model is generated in which the multiplicative random deposition is combined with a diffusive process to introduce spatial autocorrelation.   In each model a 2D random field is constructed by repeated deposition on uniformly random sites on a 1000x1000 pixel lattice.  In the additive and multiplicative models, a total of 100,000,000 deposition events are simulated.  In the multiplicative-diffusive model deposition occurs in 1000 cycles of progressively increasing deposition in which each deposition cycle is coupled with a diffusion process simulated by convolution with a 3x3 Gaussian kernel.  As predicted by the Central Limit Theorem, the resulting distributions of 1,000,000 pixel values have a normal distribution ($\mu=100$, $\sigma=10$) for the additive model and lognormal distributions for the multiplicative ($\mu=4.9$, $\sigma=0.49$) and multiplicative-diffusive ($\mu=17.8$, $\sigma=1.3$) models.  Progressive segmentation was applied to each 2D random field generated by the models and rank-size distributions were computed from the resulting segments.   The diminishing segmentation thresholds were chosen above and below the medians of the grid distributions where the network structures evolve most rapidly.

# Results

*Global Scaling of Settlements, Agriculture and Forests*

Settlements, agriculture and forests all show similar scaling for a variety of products and range of thresholds. Rank-size distributions are compared for each land cover type in Figure 3. Both night lights and ambient population density have slopes near -1 for multiple thresholds spanning a factor of four in brightness and an order of magnitude (10 to 100 people/km$^2$) in population density. The highest density threshold of 1000 people/km$^2$ yields a lower slope of -0.78, which is not surprising since it is high enough to fragment medium density peri-urban settlement networks in most parts of the world. As expected, both night light brightness and ambient population density distributions increase in slope as thresholds approach median values and network components begin to grow and connect more rapidly thereby shifting the mass of the distributions toward the upper tails. For both products, the range of segment sizes that is well fit by the power law spans more than three orders of magnitude.

Global rank-size distributions of contiguous areas of agricultural land cover have consistent slopes over four orders of magnitude difference in size. Power law fits to each rank size distribution yields similar exponents near -1 for the IIASA-IFPRI and MODIS products for both 25% and 50% cultivation thresholds. The slopes of the Hyde and Earthstat products are higher for both thresholds. The larger cutoffs of these products result from the lower 10x10 km resolution and greater quantization of smaller segments in the lower tails. The small cutoffs for the IIASA-IFPRI and MODIS products indicate that the power law fits all but the smallest of the areas in the distributions.

Global forest cover distributions have slopes slightly steeper than -1 for thresholds ranging from 40% to 80%. At regional scales, forest cover distributions derived from the intermediate threshold of 60% have more consistent slopes spanning -1 for all regions except Oceania. The exception of Oceania is not surprising since the size distribution of forests is constrained by the size distribution of the islands containing the forests. Apparently the process controlling the size of islands does not yield a slope of -1, although the distribution is long tailed and might still be reasonably described as a power law.

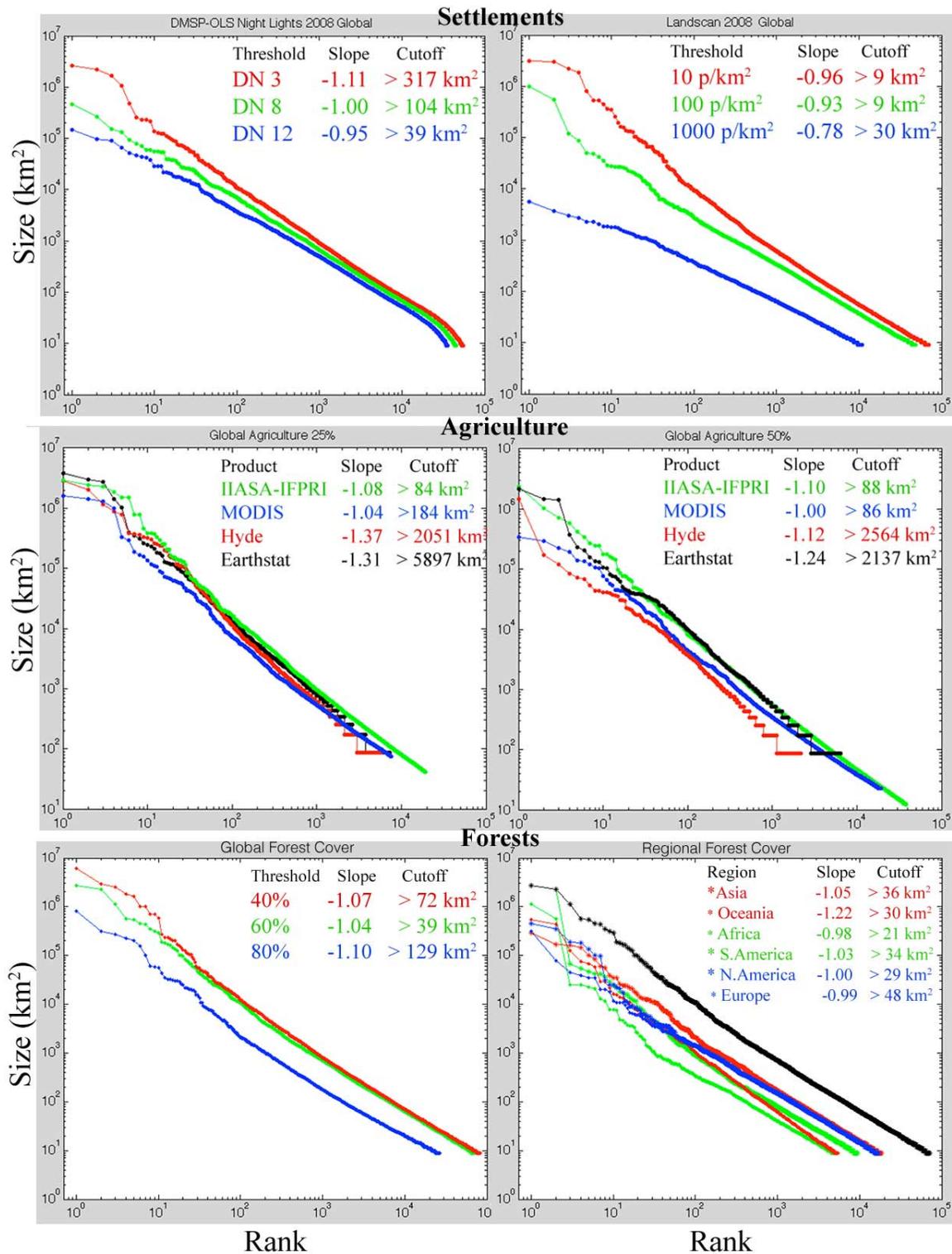

Figure 3  Spatial scaling of global land cover.  Rank-size distributions of spatially contiguous segments of land cover show consistent slopes over 4+ orders of magnitude for a wide range of thresholds for all products shown in Fig. 1.  Power law fits to each distribution yield slope exponents near -1.  Size cutoffs give lower bound of power law behavior based on optimization of a goodness-of-fit statistic.  Slopes near -1 indicate that land cover segments diminish in size at the same rate that they increase in number.

*Spatiotemporal Evolution of Settlement Networks in China and India*

Tri-temporal network growth maps and rank-size distributions for China and India in 1994, 2004 and 2013 are shown in Figure 4. The network growth maps show $Log_{10}$ segment sizes as varying brightness of blue, green and red for 1994, 2004 and 2013 respectively. Brighter colors correspond to larger segments. Warmer colors (e.g. yellow, red) indicate growth over time while cooler colors (e.g. cyan, blue) indicate shrinkage. White areas are contained within the largest networks in all three years. Because the greater brightness of the largest segments have much greater contrast against the black background, the tri-temporal network growth maps emphasize the largest network components while the smaller components corresponding to isolated settlements tend to fade into the background.

The network growth maps clearly show the establishment of several regional networks of lighted development in both China and India. In China the largest components are on the Yangtze and Pearl River deltas and the North China Plain. Each component is composed of cities spanning a range of sizes, connected by a network of smaller settlements distributed more evenly over the agricultural matrix of each basin. In India the largest components of the network form corridors between major cities with less obvious geographic correspondence than is apparent in China. Aside from the New Delhi – Lahore corridor on the North Ganges Plain, the other corridors are not associated with agricultural basins (although agriculture is pervasive throughout India). A comparative analysis of the geography of India and China is beyond the scope of this study, but the differences in the spatial structure and growth of their respective networks of development highlight differences in their trajectories of development over the past 20 years.

The rank-size distributions of lighted development in China and India illustrate contrasting forms of spatial network growth. Whereas India shows pronounced growth of the eight largest network components between 2004 and 2013, with more modest growth of the smaller components, the entire distribution of network components increases in size over both time intervals for China. The three largest Chinese networks are approximately the same size in 1994, but now span almost an order of magnitude difference in size as the North China Plain component has expanded southward from Beijing to encompass the network of smaller cities on the plain and the corridor of larger cities between Beijing and Xi'an. In 2004 the Yangtze Delta component briefly surpassed the North China Plain component in size, but by 2013 the latter had become much larger. In contrast, the New Delhi – Lahore corridor is, by far, the largest component in the Indian network in all three years. The Coimbatore-Chennai corridor is the second largest in all three years, but not by a substantial margin. The Hyderabad-Rajahmundry corridor has established itself only recently, growing by more than an order of magnitude in area since 2004. Note that the nucleation of new lighted settlements at the bottom of the distributions has steadily increased in number for both India and China. All the rank-size distributions for night lights roll off in their lowermost tails because of varying detection limits for the smallest, dimmest settlements approaching the resolution limit of the OLS sensor.

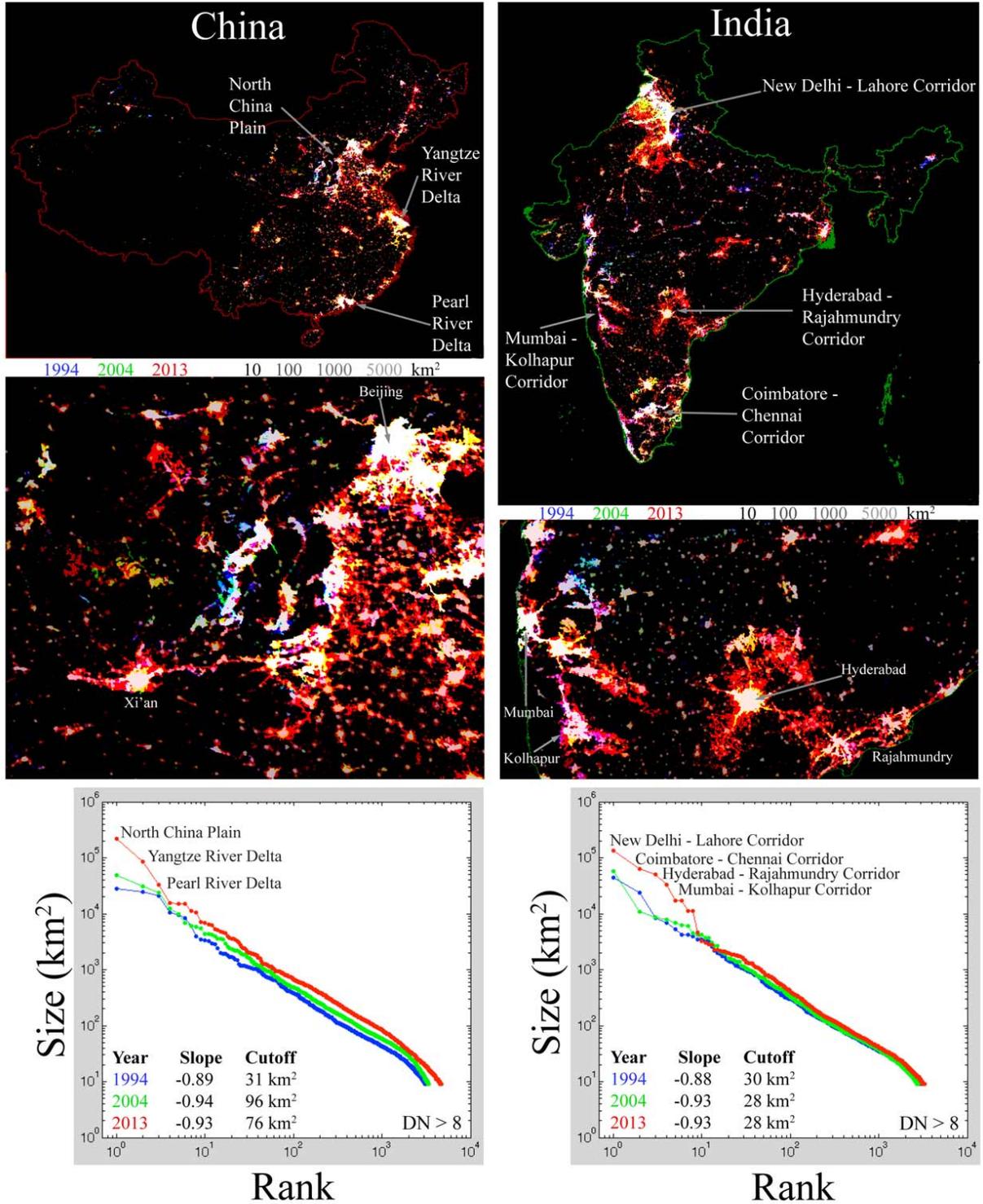

Figure 4  Growth of spatial networks of night light in China and India.  Brightness of blue, green and red in network growth maps (top, middle) correspond to Log$_{10}$ segment area in 1994, 2004 and 2013 respectively.  Warmer colors indicate growth and interconnection.  Rank-size distributions (bottom) maintain slopes somewhat less than -1 with slight increases in slope since 1994.   Parallel upward progression suggests that components are nucleating, growing and linking at comparable rates.

*Random Spatial Network Evolution and Scaling*

The progression of spatial networks and distributions that result from the progressive segmentation of the random spatial fields have several similarities – both to each other, and to the spatial networks of land cover described above. We illustrate these similarities using the random additive model and the random multiplicative-diffusive model. The results of the random multiplicative model are encompassed by these two models. Starting in the upper tail of each random field distribution of accumulated depositions, we progressively segment the grid with decreasing thresholds to simulate the growth of a spatial network of areas above threshold against a background of areas below threshold. This progression is illustrated for both random network models in Figure 5. We refer to the spatially contiguous areas above threshold as the components of the foreground network and those below threshold as the background network. The components of the foreground network are shown on a succession of network maps, color coded by area with warmer colors corresponding to larger components with the largest component shown white and the entire background below threshold shown black. The rank-size distributions of the foreground and background network components are shown together on the same plot for each successive threshold. In addition, logarithmically-binned histograms of the $Log_{10}$ of total area in each size interval are inset to show the total area occupied by each component size range within the domain.

The progressive evolution of both random network models are first described in narrative to illustrate similarities, then several general observations are drawn in comparison to the land cover networks described above.

*Phase 1 – Nucleation Dominant*
Initially, at high thresholds, only the pixels in the uppermost tails of the random field distributions (shown inset in the uppermost network maps in Figure 5) exceed threshold. These pixels are the isolated nuclei that will eventually grow into network components. As the thresholds are decreased, more nuclei emerge and some pixels exceed threshold adjacent to previous nuclei. These adjacent sets of pixels form the first multi-pixel components. The uppermost network maps and distributions illustrate the dispersed nature and small range of sizes of these early components. At this stage, the background network is fully connected as a single giant component with a size only slightly smaller than the total domain area. Even at this early stage, there are several orders of magnitude more small (1-5 pixel) components than larger (10-100 pixel) components so the rank-size plots have very little slope.

*Phase 2 – Growth Dominant*
As the thresholds are decreased further, more adjacent pixels exceed threshold and all components begin to grow faster. The curvature and slope of the foreground network rank-size distributions continue to increase as larger networks form by connection in the upper tail and more nuclei emerge in the lower tail. The total number of components (maximum rank) increases. The giant component of the background network abruptly disappears in the additive model but in the multiplicative-diffusive model it becomes fragmented and a distribution of component sizes begins to emerge (thin curves) for the background network. As the threshold begins to ascend from the upper tail to the mode of the random field distribution, the rate of exceedance increases. Preferential attachment in the foreground network accelerates as the largest components grow increasingly faster in size. The curvature of the rank-size distributions increases as larger components grow and connect faster than smaller components. The total number of components begins to decrease as the rate of component connection exceeds the rate of nucleation. The combination of growth and connection pushes the center of the rank-size distribution upward and leftward toward the upper tail as the lower tail shrinks and becomes steeper. The range of individual component sizes now spans three orders of magnitude but the range of total areas of all component size intervals spans less than one order of magnitude (inset histograms). The distribution of total component area becomes increasingly uniform over an increasingly wider range of component sizes.

*Phase 3 – Connection Dominant*
As the background network becomes increasingly fragmented by the growth and connection of the foreground network, its rank-size distribution becomes more similar to that of the foreground network – until their trajectories cross. The giant component of the background network is fragmented and disappears relatively early in the random additive model but persists longer in the random multiplicative-diffusive model because of the autocorrelation induced by the diffusion of the foreground network. Eventually, the rate of connection of components in the foreground network exceeds the combined rates of growth and nucleation as the threshold approaches the median of the distribution of the random field. The result is cannibalization of smaller components by connection to form larger components and the total number of components diminishes more rapidly – collapsing the rank-size distribution toward the upper tail while increasing its slope and reducing its curvature. This gives rise to a rank-size distribution with minimum curvature and maximum slope. Now the growth of the largest components can no longer be accommodated by the domain and they begin to connect to each other. The total area of components is nearly equal in each size interval and the histograms show a nearly uniform distribution with size over a range of four orders of magnitude. The slope of the rank-size distribution approaches -1. As the threshold passes the median of the distribution of the random field, the largest components run out of space and are forced to connect to each other. Explosive percolation is inevitable and the foreground network components coalesce into a single giant component. The symmetry of foreground and background components becomes apparent as their respective rank-size distributions resemble their complements at the beginning of the process.

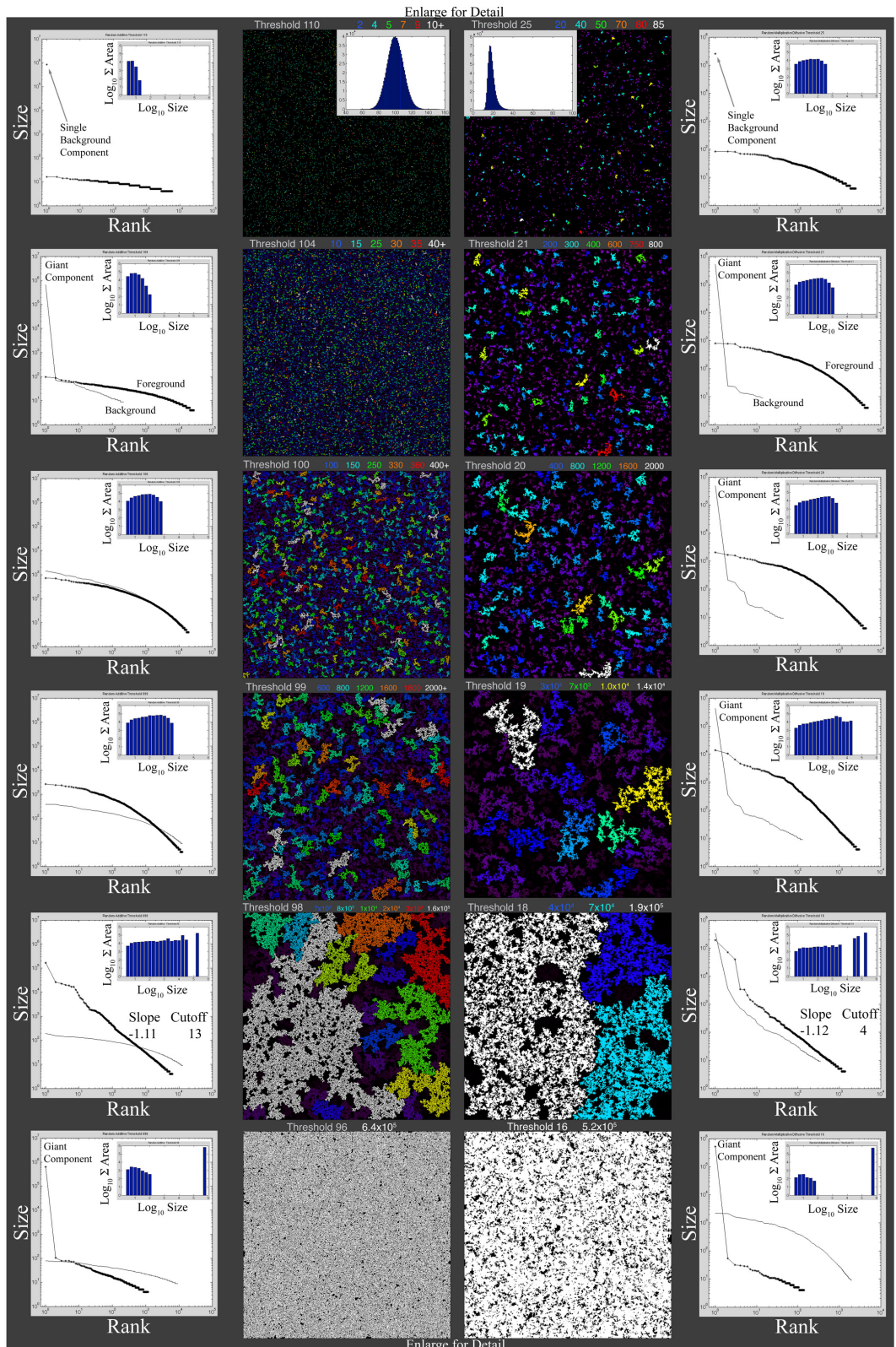

Figure 5  Growth of spatial networks by progressive segmentation of continuous random fields. Both the uncorrelated additive model (left) and diffusive multiplicative model (right) show similar progressions of network formation and component size distributions from nucleation (top) to growth and linking, finally culminating with explosive percolation (bottom).  Background complement networks behave similarly.

The random field models illustrate several fundamental characteristics of bounded spatial networks that are also observed in the global land cover networks. The progression of network structure and component size distributions displayed by the progressive segmentation of the random fields reveals several general properties of bounded spatial networks.

*Genesis of nodes as a consequence of progressive segmentation over a field distribution*
Progressive segmentation provides a mechanism to characterize the spatial structure of a continuous field as a progression of bounded spatial networks. The structure and evolution of the networks are determined by the spatial structure and distribution of field values. Genesis of nodes, and the three processes that result, are analogous to temporal evolution of a system in which the threshold corresponds to time and the evolution is prescribed by the field values. This allows temporal evolution to be represented as spatial structure.

*Nucleation, growth and linking as consequences of node genesis*
The relative rates of these three fundamental processes are determined by the spatial structure of the field and determine the evolution of the network.

*Preferential attachment as a consequence of genesis on a bounded domain*
The process of preferential attachment, giving rise to scale-free networks, occurs naturally on a bounded spatial network because potential attachment is proportional to perimeter and larger components' perimeters occupy proportionally more of the limited area of the domain.

*Linking of components as a consequence of preferential attachment*
Larger components grow increasingly faster because their larger perimeters offer more opportunity for connection to other components. Connection offers the greatest opportunity for abrupt growth because it scales with the sum of the sizes of the components relative to the total area of the domain.

*Reduction of components as a consequence of linking*
The total number of components in the network is reduced, and the size of the resultant components increases whenever components connect. Connection results in the formation of the giant component and the collapse of the rank size distribution when explosive percolation occurs.

*Co-evolution of complementary networks on a bounded domain*
Evolution of foreground networks control evolution of background networks on bounded domains. Co-evolution of both is determined by the spatial structure and distribution of the continuous field.

**Discussion**

The simulations show that even spatial networks derived from progressive segmentation of purely random fields can yield rank-size distributions with upper tails approaching a unity power law with a slope of -1. This supports our assertion that the scaling properties and uniform area distribution implied by a unity power law rank-size distribution may be a general characteristic of bounded spatial networks produced by segmenting certain continuous fields. The implication is that process-specific mechanisms may not be required to explain the rank-size distributions observed for the diversity of land cover products described here.

The persistence of the unity power law raises the question of what is special about a power law with a slope of -1. The random field models attain slopes of -1 along a progression of rank-size distributions, most of which have different slopes and greater curvature. Unless it is just a coincidence that we observe all of these land cover distributions at the same point along a progression of configurations, the observation seems to suggest some form of stability. Perhaps the most persuasive evidence for this assertion is the persistence of the near unity slope of the growing distributions of Chinese and Indian night lights. However, the suggestion that the unity power law distribution may represent a stable network configuration seems particularly paradoxical given its emergence immediately preceding explosive percolation in the random field simulations.

The persistence of the unity slope has implications for the evolution of the network. A power law rank-size distribution with a slope of -1 implies that the network components increase in number at the same rate that they diminish in size. This implies a uniform distribution of total area across all sizes of component – as seen in the random field simulations. In a growing network, this implies that the rates of nucleation, growth and connection must balance to maintain the unity slope. If the rate of nucleation outpaces the rate of growth the network becomes dominated by small components and the slope of the rank-size distribution decreases. If the rate of growth outpaces the rate of nucleation the network becomes dominated by intermediate size components and the curvature of the rank-size distribution increases while the slope of the lower tail increases and the slope of the upper tail decreases. If the rate of connection outpaces the rate of growth, the network becomes dominated by large components and the slope and curvature of the upper tail increases. Each of these situations occurs at some point in the evolution of both random field network models. At one point, however, the rates are balanced and a unity power law emerges – immediately before explosive percolation occurs.

We conjecture that the persistence of unity power law rank-size distributions in different types of land cover suggests the existence of negative feedback that prevents explosive percolation of land cover networks. The geographic distribution of conditions conducive to settlements, agriculture and forests obviously imposes a fundamental constraint. However, even within areas conducive to all of these land cover types, we observe interspersed networks of settlement, agriculture and forest. In these areas, the forest that preceded development might be considered the background network and the agriculture and settlements could be considered foreground networks. Speculation about the driving forces of land cover change and their potential roles as feedbacks is beyond the scope of the present study.  Nonetheless, the persistence of unity power law rank-size distributions spanning 4 to 5 orders of magnitude in size and number provides a powerful constraint for testing theories and models of land cover dynamics.

The unity power law rank-size distribution is not a general property of all spatial networks derived from progressive segmentation of continuous fields.  We have found one prominent exception so far.  Applying progressive segmentation to digital elevation models (DEMs) of Earth's continents does not generally yield rank-size distributions with power law upper tails or slopes near -1.  Progressive segmentation of 500 m resolution continental DEMs from the USGS yields rank-size distributions with discontinuous upper tails and greater curvature than is observed for the land cover products.  Furthermore, it is necessary to use different ranges of elevation thresholds for each continent to obtain long-tailed rank size distributions bearing any resemblance to those observed for the land cover products.  The slopes ($\mu \pm \sigma$) we obtain from progressive segmentation of DEMs for Asia, Australia, Europe+Africa, North America and South America are -1.49 ± 0.2, -1.43 ± 0.09, -1.48 ± 0.09, -1.39 ± 0.2 and -1.39 ± 0.12 respectively.

**Acknowledgements**
Much of the research summarized here was funded by NASA, and by the NASA Land Cover and Land Use Change (LCLUC) program in particular. CS was funded most recently by the NASA LCLUC program (grant LCLUC09-1-0023) and Interdisciplinary Science program  (grants NNX12AM89G & NNN13D876T) and by the NASA Socioeconomic Data and Applications Center (SEDAC) (contract NNG13HQ04C). Earlier related funding to CS was provided by the National Science Foundation (grant DEB 09-48451), the US Dept. of Agriculture (grant NYR-2006-01697), the United States National Institutes of Child Health and Development (award R21 HD054846) the UCAR Visiting Scientist Program and the Doherty Foundation.  Many of the perspectives presented here have evolved from years of discussions with Deborah Balk, Ligia Barrozo, Uwe Deichmann, Chris Elvidge, Geoff Henebry, Cristina Milesi, Mark Montgomery, Reinaldo Peréz-Machado, Son Nghiem, Francesca Pozzi and Greg Yetman. DS is supported by a National Defense Science and Engineering Graduate Fellowship (NDSEG) through the U.S. Department of Defense. DS thanks L. Sousa for her love and support.